\documentclass[aps,preprint]{revtex4}
\usepackage{graphicx}
\usepackage[dvips]{epsfig}
\begin{document}
\title{Effects of disorder on the Raman line-shape  in ZrO$_2$}
\author{L. A. Falkovsky \footnote{e-mail: falk@itp.ac.ru}}
\affiliation{Landau Institute for Theoretical Physics, 119337
Moscow, Russia}
\begin{abstract}
Experimental data obtained by Lughi and  Clarke  [J. Am. Ceram.
Soc., to be published] are compared to the theory describing
disorder effects on  optical phonons. Sharpening and  vanishing of
the asymmetry of the Raman lines after annealing are attributed to a
decrease in short-range disorder. The parameters of disorder (such
as the phonon-defect coupling and the correlation length) are
defined in the comparison.
\end{abstract}


\maketitle

\section{Introduction}
 Raman scattering has been used for  years  to measure the
center-zone phonon frequencies. But  it was understood quite
recently that  studies of the  Raman line shape  can also provide
widespread information relative to disorder, for instance,  the
isotopic compositions \cite{GRZ}, the stacking faults
\cite{NNH}-\cite{RHL}, the strain fluctuations near interfaces
\cite{FBC}, and so on. The phonon scattering processes by defects
of various types result in increasing the phonon relaxation rate and
consequently   the Raman line width.

A novel field of Raman spectroscopy,  the transformation
kinetics, was put forward  by  Lughi and  Clarke \cite{LC}. They
investigated the phase transformations of yttria-stabilized
zirconia.Zirconia ZnO$_2$  has the wide range of applications,
including traditional structural refractories, fuel cells, and
electronic devices such as oxygen sensors. Because of these
technological implications, the characterization of zirconia is of
particular interest. Zirconia has three zero-pressure polymorphs:
the high-temperature cubic ($c$) phase (stable between 2570 K and
the melting temperature  2980 K), the tetragonal ($t$) phase
(stable between 1400  and 2570 K), and monoclinic ($m$) phase
below 1400 K.

The unit cell of the $c$ phase contains one formula unit. There is
only one triply degenerated Raman active mode at 640 cm$^{-1}$
\cite{AB},\cite{CRR}. From the symmetry analysis,
 it is suggested \cite{ID}-\cite{NT} that the $c-t$ transformation
of ZnO$_2$ results from a condensation of the Brillouin zone
boundary phonons. Indeed, it was found in the model calculation
\cite{MSQ},\cite{FPF} of phonon curves, that the phonon frequency
$\omega(X_2^{-})$  of the $c-$phase is close to zero. The volume
of the elementary cell is doubled in the $c-t$ transformation. In
the $t$ phase, there are six Raman active modes \cite{APA},
\cite{BGL} (their frequencies are given in Table).

In the work \cite{LC}, zirconia coatings were obtained by electron
beam evaporation and then annealed. The X-ray diffraction and
Raman spectroscopy measurements were performed before and after
annealing. They give a possibility to control the phase
transformation in the samples. The obtained results are very
interesting because they demonstrate the gigantic  effect of
annealing on both the width and asymmetry of the Raman line.

 In the present paper we compare the results  \cite{LC} with the
theory \cite{Fan} describing the  disorder effect on the optical
phonons. In Sec. II, a sketch of the theory is presented. The
comparison  of the theory with the experiment \cite{LC} are given in
Sec. III. In Sec. IV,  concluding remarks are summarized.

\section{Theoretical background: phonon elastic  scattering by static disorder}

The scattering of phonons by disorder (defects or strain
fluctuations) results in the phonon width and shift (see \cite{usp} for
review ). To estimate the effect of disorder
we can use the  quantum-mechanical  formula \cite{Tam}:
\begin{equation}\label{qm}
\Gamma(\omega,{\bf q})-\Gamma^{\rm nat}\propto \pi c_d \sum_k
|v({\bf k} - {\bf q})|^2 \delta(\omega_0^2\pm s^2k^2-\omega^2),
\end{equation}
giving the phonon width as a function of the frequency and wave
vector. Here $c_d$ is the concentration of defects, $v{(\bf k} -
{\bf q})$ is the phonon-defect interaction, $\omega$ is the phonon
frequency in the initial state, $\omega_0^2\pm s^2k^2$ is the
squared phonon frequency in the final state (the "$+$"\, sign if
 the phonon branch has a minimum at $\omega_0$ and
 the "$-$"\, sign for the maximum), and
$\delta(x)$ is the Dirac $\delta$-function expressing the
conservation low in the phonon elastic scattering. The summation
is performed over the 3-dimensional wave vector ${\bf k}$ in the
case  of phonon scattering by the point
defects under consideration.

Let us consider  the short-range defects with the interaction having
the form of the $\delta$-function in the real space. Then, the
potential $v$ is independent of the wave vector, i.e. $v({\bf k} -
{\bf q})=v_0$ with a constant value $v_0$, and  the integration
can be  done explicitly. For the phonon minimum,  the contribution
of the scattering by defects appears only in the region
$\omega^2>\omega_0^2$:
\begin{equation}\label{est}
\Gamma(\omega)=\Gamma^{\rm{nat}}+\frac{ c_d}{4\pi \omega_0
s^3}|v_0|^2 \sqrt{\omega^2-\omega_0^2}.
\end{equation}
In the case   of the phonon branch maximum, we have the disorder
contribution  only if $\omega^2<\omega_0^2$.

Expression (\ref{est}) has a transparent physical meaning: the
phonon elastic  scattering from defects produces some phonon width
in that frequency range,  where  the appropriate final
phonon states occur. For instance, in the case of the minimum of the
phonon branch at $\omega=\omega_0$, the final phonon states are
only for $\omega>\omega_0$. Of course, this effect manifests
itself as a line asymmetry in the Raman light scattering on
phonons. As a result, the high-frequency wing of the line drops
more slowly (than the low-frequency one) for a phonon branch
having the minimum at the center of the Brillouin zone.

I should  emphasize that the frequency dependence of the
phonon-defect width is determined by the dimensionality of
defects. As we see from Eq. (\ref{est}), this is a square-root
dependence for  point defects. In the case of linear
defects
 such as dislocations (or the plane defects -- crystallite
 boundaries or stacking faults), we have the 2d (or 1d)
 integration in Eq. (\ref{qm}) which gives
\begin{eqnarray}  \label{2}
 \Gamma(\omega)=\Gamma^{\rm{nat}}+\frac{ c_l}{4\omega_0 s^2}|v_0|^2
\theta(\omega^2-\omega_0^2),\,{\rm  line \, defect,}\\
\label{3}
 \Gamma(\omega)=\Gamma^{\rm{nat}}+\frac{ c_p}{4\omega_0 s }|v_0|^2
(\omega^2-\omega_0^2)^{-1/2},\, {\rm plane \, defect},
\end{eqnarray}
where $\theta(x)$ is the Heaviside step-function.

The singularities  in Eqs. (\ref{2}) - (\ref{3}) as well as the
weak singularity in Eq. (\ref{est}) arise, because the width of
final states was not taken into account. There is another
shortcoming of the simple expressions (\ref{qm})-(\ref{3}): they
do not describe the shift of the phonon frequency due to the
interaction with disorder. Therefore, we have to use a more
complicated technique of the Green functions \cite{Fan}.

The heart of the technique is the phonon self-energy
\begin{equation}
\label{se} \Sigma(\omega)=-c_d\sum_{k } \frac{ |v({\bf k} - {\bf
q})|^2} { \Omega ^{2}(\omega) \pm s^{2}k^{2} - i\omega  \Gamma
(\omega)- \omega ^{2} }.
\end{equation}
The  real functions $\Omega (\omega)$ and $\Gamma (\omega)$
are  now themselves defined by the real and imaginary parts of the
self-energy. We obtain them by solving  the system of the equations
\begin{equation}
 \label{main}
\Omega ^{2}(\omega)- \omega _{0}^{2} - i \omega [\Gamma
(\omega)-\Gamma^{\rm nat} ] - c_dv(q=0) =  \Sigma(\omega).
\end{equation}
The last term (named  the homogenous shift) in the left-hand
side of Eq. (\ref{main}) gives the phonon shift due to the
averaged effect of disorder,
 whereas the self-energy, Eq. (\ref{se}), describes the fluctuation effect of the
 defects.
 In addition to the phonon shift (inhomogeneous, depending on the frequency),
 the self-energy produces  a phonon
  width. Note, that
if the phonon does not interact with disorder ($v=0$), 
Eqs. (\ref{main}) naturally give $\Omega (\omega)=\omega_0$
and $\Gamma (\omega)=\Gamma^{\rm nat}$. Next, in the Born
approximation, Eqs. (\ref{se})-(\ref{main}) give the simple
expression, Eq. (\ref{qm}), for the phonon damping.

In the  Raman light scattering,  the wave-vector transfer
$q\sim\omega^{(i)}/c$ is determined by the wave vector of incident
light, $\omega^{(i)}$ being the incident light frequency. In the
integral (\ref{se}), the values of $k\sim 1/l=\sqrt{\Gamma\omega_0}/s$
are important, where $l$ plays the role of the phonon mean free
path. The dispersion parameter $s\sim 5\times10^5$ cm/s
 is typically of the order of the sound velocity and the  width $\Gamma$
 is of the order of $(10^{-1}\div 10^{-2})\omega_0$.
Thus, the condition $q\ll k$ holds and we can omit $q$ in Eq.
(\ref{se}).

To simplify the calculations, we assume that the
potential function $v({\bf k})$ takes  constant value $v_0$ in
the region $k<1/r_0$ and  is equal to zero for $k>1/r_0$. Then
the parameter $r_0$ gives the size of the region in the real
space, where phonons are scattered by defects. If we consider the
phonon scattering by  strain fluctuations, $r_0$ has the meaning
of the correlation length of these fluctuations.

The line asymmetry is very sensitive to the relation between the
parameters $r_0$ and $l$. The case where $r_0<l$ is referred  to as the
short-range disorder. It is easy to verify that the line asymmetry
is much larger for the short-range disorder than in the opposite
case. We can rewrite the condition of the short-range disorder as
\begin{equation}\label{sr}
\pi \frac{r_0\sqrt{\Gamma\omega_0}}{a\omega_D}<1,
\end{equation}
  using the
estimate of the Debye frequency $\omega_D=\pi s/a$, where $a$
is the lattice constant.

For the point defects, the calculation of $\Sigma(\omega)$, Eq.
(\ref{se}), gives
\begin{eqnarray}
\label{sgma} \Sigma(\omega)=A \Biggl[ b - (a_1 - i a_2) \Bigl(
\frac{1}{4}
 \ln \frac{(b +a_1)^2+a_2^2}{(b -a_1)^2+a_2^2} \\
+ \frac{i}{2} \arctan \frac{b +a_1}{a_2} + \frac{i}{2}
\arctan\frac{b -a_1}{a_2}\Bigr) \Biggr] \nonumber
\end{eqnarray}
in the case of  maximum of the phonon branch and
\begin{eqnarray}
\label{sigma} \Sigma(\omega)=A \Biggl[ -b + (a_2 + i a_1) \Bigl(
\frac{1}{4}
 \ln \frac{(b +a_2)^2+a_1^2}{(b -a_2)^2+a_1^2} \\
 -\frac{i}{2} \arctan \frac{b +a_2}{a_1} -\frac{i}{2}
 \arctan \frac{b -a_2}{a_1}\Bigr) \Biggr] \nonumber
\end{eqnarray}
in the case of  minimum. Here
\begin{eqnarray}\nonumber
a_1=[\Omega^2(\omega)-\omega^2 + \rho(\omega)]^{1/2}/\sqrt2, \\
\nonumber
 a_2=  [-\Omega^2(\omega)+\omega^2 +
\rho(\omega)]^{1/2}/\sqrt2,\\ \nonumber
\rho(\omega)=[(\Omega^2(\omega)-\omega^2)^2+
\omega^2\Gamma^2(\omega)]^{1/2},\\ \nonumber
 b= s/r_o,\quad A =c_dv_0^2 /2 \pi^2 s^3 .
\end{eqnarray}
 Notice, that  the dimensions are  the following: $s$ [cm
$\omega$], $c_d$ [1/cm$^{3}]$, $v_0$ [cm$^{3}\omega^2$],   $A$
[$\omega$].

Solving the system of Eqs. (\ref{se})-(\ref{main}), we find the
functions $\Omega (\omega)$ and $\Gamma (\omega)$. Then we can
calculate (see, i.e., \cite{FBC}) the Raman intensity
\begin{equation}\label{int}
I(\omega)\sim
\frac{\omega\Gamma(\omega)}
{[1 - \exp(-\hbar\omega /k_B T)]
[(\Omega^2(\omega)-\omega^2)^2+\omega^2\Gamma^2(\omega)]}.
\end{equation}
  Equation (\ref{int})
can be applied to the Stokes lines ($\omega>0$), as well as to the
anti-Stokes ones ($\omega<0$).

If $\Omega (\omega)=\omega_0$ and $\Gamma (\omega)=\Gamma^{\rm
nat}$ are constant, Eq. (\ref{int}) gives the Lorentzian
line-shape, because $\Gamma^{\rm nat}\ll\omega_0$. But, as we have
already seen,
 the non-Lorentzian line-shape is obtained if the width
$\Gamma(\omega)$ depends on $\omega$.

\begin{figure}[h]
\begin{center}
\resizebox{.50\textwidth}{!}{\includegraphics{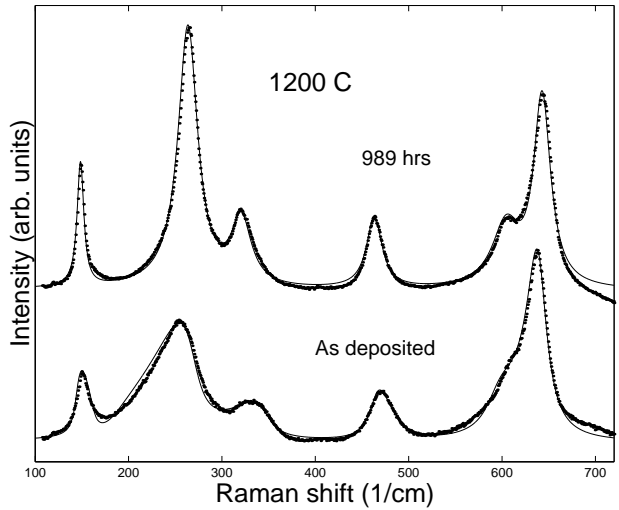}}
\end{center}
\caption{ Raman spectra (points - the experiment data, solid lines
- the theory) from the sample in as-deposited conditions (bottom)
and after annealing for 989 hrs at 1200 $^o$C (top). \label{rf1}}
\end{figure}

\begin{figure}
\begin{center}
\resizebox{.50\textwidth}{!}{\includegraphics{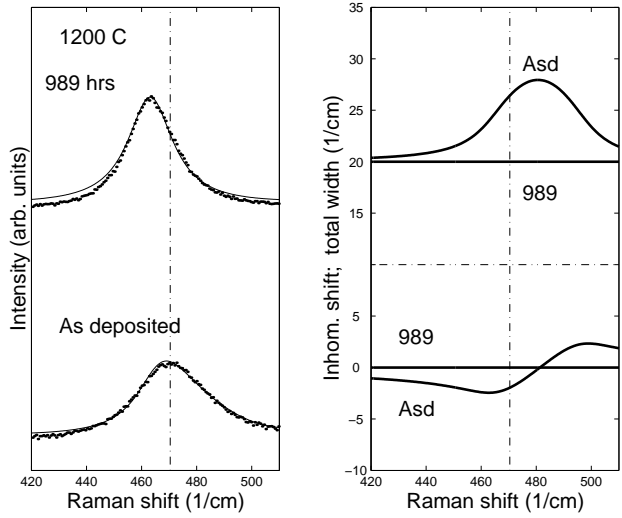}}
\end{center}
\caption{Raman line 470 cm$^{-1}$ (points - the experiment data,
solid lines - the theory) from the sample in as-deposited
conditions (left panel, bottom) and after annealing for 989 hrs at
1200 $^o$C (left panel, top); the peak of line from the
as-deposited sample is marked by the vertical dash-dotted lines.
The asymmetry of the line indicates that the phonon branch has a
minimum.
 In the right panel, the total width $\Gamma(\omega)$ (top)  and the
inhomogeneous shift $\Omega(\omega)-\omega_0$ (bottom) are plotted
versus the frequency transfer for two spectra.} \label{rf2}
\end{figure}

\begin{figure}
\begin{center}
\resizebox{.50\textwidth}{!}{\includegraphics{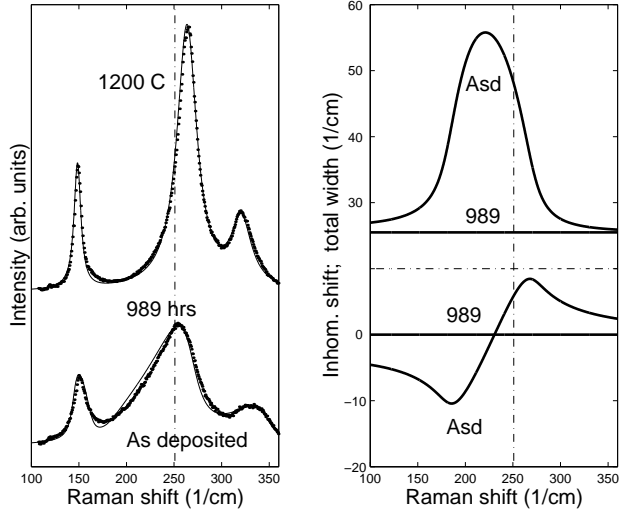}}
\end{center}
\caption{ Same as Fig. 2, but for the line 250 cm$^{-1}$. The
phonon branch now has a maximum.} \label{rf3}
\end{figure}

Up to this point we have considered a single phonon line. In the
general case there are several lines  with their positions
$\Omega_i(\omega)$ and widths $\Gamma_i(\omega)$. The corresponding
terms have to be summarized in Eq. (\ref{int}). The intervalley
terms with $v_{ij}$ could be written also in the self-energy
(\ref{se}). These terms are essential  when the disorder removes
the  degeneration of  branches.

\section{Comparison of the theory with the experimental data for zirconia}

Two Raman spectra taken from  paper \cite{LC} are shown in Fig. 1.
They were recorded from  two samples of yttria-stabilized (8.6
mol\% YO$_{1.5}$) zirconia -- the first in as-deposited condition
(bottom) and the second (top) annealed  at 1200 $^o$C.  Narrowing
and  decrease of asymmetry of the lines are evident. The fit to
the theory, Eqs. (\ref{sgma})-(\ref{sigma}), is shown in solid
lines; parameters of the fit are listed in Table. In fitting, we
assume that the effect of disorder on the spectra from the sample
after annealing is negligible ($A=0$). There is no information
about the width of Raman lines for pure zirconia. Therefore, the
line widths for the sample after annealing are referred to as the
natural widths $\Gamma^{\rm nat}$.

 The broadening $\Gamma-\Gamma^{\rm nat}$ of the line
for the sample in  as-deposited conditions is induced by the
phonon-disorder  interaction determined by the parameter $A$. We
assume, that the natural width $\Gamma^{\rm nat}$ of each line
before annealing is equal to $\Gamma^{\rm nat}$ of the same line
after annealing (see Table).

The line shape of  spectra for the sample with disorder is
determined by two factors: (i) the value of the parameter $r_0$
and (ii) the extremum type    (minimum or maximum) of the phonon
branch. In Figs. 2 and 3 the results of fitting are shown in
detail.  The  470.5 cm$^{-1}$ line is shown in Fig. 2, left panel.
The high-frequency wing of the line from the  as-deposited sample
is more gentle. This corresponds to the minimum of the phonon
branch. In the right panel, the width $\Gamma(\omega)$ and the
inhomogeneous shift ($\Omega(\omega)-\omega_0$) are plotted on the
top and bottom, respectively, as functions of the frequency
transfer $\omega$. For the sample after annealing, they are
constant $\Gamma(\omega)=\Gamma^{\rm nat}$ and
$\Omega(\omega)=\omega_0$.

We see that the width $\Gamma(\omega)$ of the line from
as-deposited sample increases with the frequency transfer around
the line peak  $\Omega_0$, which is marked by the dash-dotted
vertical line. This  resembles the square-root dependence  in Eq.
(\ref{est}) for the minimum of the phonon branch and we use Eq.
(\ref{sigma}) in fitting. At the peak position, the width takes
the value $\Gamma_0$; this value is given in Table, it
 is  slightly less than the FWHM (full-width at half-maximum)
 because of the frequency dependent $\Gamma(\omega)$.

The behavior of the 250 cm$^{-1}$ line (see Fig. 3) from the
as-deposited sample has the opposite character: the low-frequency
wing drops more slowly. Then we use Eq. (\ref{sgma}) corresponding
to the maximum of the phonon branch.  The asymmetry parameter $As$
 listed in Table is a difference between the right and left
wings at the half-maximum. This parameter takes positive
(negative) values for  the phonon-branch minimum
(maximum). The asymmetry parameter and the FWHM were
determined for the line extracted from the total spectra with the
help of the fit.

Several points can be noted in Table. First, the radius of defects
(correlation radius) is about  the value of the lattice parameter
$a$. This is the case of a short-range disorder resulting in the
evident asymmetry of the line shape. Second, because of 
disorder, the line position is slightly shifted (by less than 5 \%),
but the line width increases considerably. This can be understood
with the help of Figs. 2 and 3 (right panels). The frequency
dependence of the shift function $\Omega(\omega)-\omega_0$ has a
zero about the line peak, whereas the width function
$\Gamma(\omega)$ takes almost its maximum value.
\begin{table}
\caption{Final values obtained in  fitting for  spectra from the
samples  after annealing at 1200 $^o$C (Ann) and as deposited
(Asd): the experimental (according to \cite{BGL}) phonon frequency
$\Omega_{exp}$, the peak position $\Omega_0$, the peak position
$\omega_0$ including only the homogeneous shift, the total line
width $\Gamma_0$ (subscript 0 is referred to as the peak position),
the natural line width $\Gamma^{\rm nat}$, the line asymmetry
$As$, the interaction parameter $A$ (all in cm$^{-1}$),
 the defect radius $r_0/a$ (dimensionless),
 the relative intensity  $I_0$, and the type of extremum.}
 \textwidth 160mm
        \begin{center}
                \begin{tabular}{|c|c|c|c|c|c|c|c|c|c|c|}
\hline \hline
 line & \quad$\Omega_{exp}$  \qquad & \quad$\Omega_0$  \qquad& \quad$\omega_0$ \qquad& \quad$\Gamma_0$ \qquad& \quad$\Gamma^{\rm nat}$
 \qquad& \quad$As$ \qquad& \quad$ A $ \qquad& \quad$r_0/a $ \qquad& \quad$I_0$ \qquad& \quad extr \qquad\\
\hline
1 Ann &149& 148.6 & 148.6  & 8.7   & 8.7   & 0    & 0   & -  & 0.215 & min\\
1 Asd &   &151.2 & 152.0  & 14.6  & 8.7   & 3.1  & 225 & 1.60& 0.19& min \\
2 Ann &269&264.0 & 264.0  & 25.5 & 25.5  & 0    & 0   & -  & 3.32& max\\
2 Asd &   &250.9 & 245.0  & 48.0 & 25.5  &$-23.7$ &225 & 0.73& 2.55& max \\
3 Ann &319&321.0 & 321.0  & 26.0  & 26.0 & 0    &  0  & -  & 1.15& max\\
3 Asd &   &330.4 & 325.0  & 45.0 & 26.0  &$-18.4$ &200 & 0.83& 1.10& max\\
4 Ann &461&463.3 & 463.3  & 20.0  & 20.0  & 0   & 0   & -  & 1.55& min \\
4 Asd &   &470.5 & 472.5  & 26.5 & 20.0   & 5.2 &100 & 1.27  & 1.45& min\\
5 Ann &602&604.5 & 604.5  & 36.0  & 36.0  & 0   & 0   & -  & 3.30& min\\
5 Asd &   &602.1 & 605.0  & 43.3 & 36.0   & 5.8 & 75  & 0.95  & 2.30 & min\\
6 Ann &648&643.0 & 643.0  & 22.0  & 22.0  & 0   & 0   & -  & 6.70 & max\\
6 Asd &   &636.4 & 633.5  & 27.2 & 22.0   &$-3.9$ &50  & 0.95  & 7.85& max\\
\hline \hline
\end{tabular}
        \end{center}
\label{tbl1}
\end{table}
\section{Discussion}

Lughi and Clarke explained their results in the following way. The
Raman line becomes more symmetric after annealing. According to
the condition (\ref{sr}), this means that the correlation radius of
disorder in as-deposited samples must be quite small,  of a few
lattice parameters. Therefore, the disorder can not be induced by
the large-scale strain fluctuations, for instance, of the
crystallite size (about 50 nm).

The disorder with small correlation length can be realized by
Y$^{3+}$ ions and their associated oxygen vacancies. But the
concentration of yttria (8.6 mol$\%$) before and after annealing
is the same in the sample. Lughi and Clarke proposed a
redistribution mechanism of the Y$^{3+}$ ions and oxygen
vacancies. They reasoned that the regions of the $c$ phase become
 richer (and the $t$ phase is more pure) in the Y$^{3+}$ ions and
oxygen vacancies in the $t-c$ phase transformation during
annealing. Because of this redistribution of defects the Raman
active $t$ phase exhibits   the symmetric and narrowing Raman
lines.

Considering the ions and vacancies as the point defects that
scatter the long-wave optical phonons, we  use  Eqs.
(\ref{sgma})-(\ref{sigma}) in our comparison.

Finally, as can be seen from Table, the phonon-disorder
interaction is larger for the low frequency lines. This
interaction decreases progressively with the frequency of the lines.
We conclude that the distribution of the heavier defect ions, i.e.,
Y$^{3+}$, affects the phonon modes stronger than the disorder of
oxygen vacancies.

\section{Acknowledgments}

I am grateful to V. Lughi and D.R. Clarke for providing their data and 
a copy of
Ref. \cite{LC} prior to publication. The work was supported by the
RFBR Grant  \# 04-02-17087.

\end{document}